\documentclass[aps,prl,preprint,groupedaddress]{revtex4-2}
\usepackage{amsmath}
\usepackage{graphicx}
\usepackage{array}
\graphicspath{ {./figures/}}
\begin{document}
	
	%\preprint{}
	\title{Entropic Analysis of Reservation Policy of Government of India}
	
	
	\author{Rakesh Kumar Pandey}
	
	\email{rkpandey@kmc.du.ac.in, r.rkr.pandey@gmail.com}
	%\thanks{}
	\affiliation{Kirori Mal College, University of Delhi, Delhi - 110007}

	\author{Maneesha Pandey}
	\email{maneeshapandey@hinducollege.ac.in, pandeymaneesha@gmail.com}
	%\thanks{}
	\affiliation{Hindu College, University of Delhi, Delhi - 110007}

	
	\date{\today}

	\begin{abstract}
		Analysis based on the Entropy estimates of the society is carried out on a mathematical model in this paper, to make an assessment of Reservation policy presently in operation in India. The Reservation Policy was provisioned in the constitution of India under the policy of affirmative action. The policy makers had identified some communities of India that were being discriminated socially. They identified them by noticing their dismal representation in positions of power and respect, within the society. It was argued that these communities are required to be treated differently to give them equal opportunity in the society. The policy makers had provisioned in the same policy that the effectiveness of this policy will be analyzed at regular intervals to decide its continuation or improvisation. An attempt is made here to scientifically analyze the effectiveness of this policy and suggest possible improvements in the same. As per the understanding of Entropy, a policy such as the Reservation Policy, that is intended to encourage equality must be associated with increase in the Entropy of the society. After a careful mathematical analysis, it is shown here that the policy, in its present form, is designed to increase the entropy of the society in the short-term scenario but is an iso-entropic policy in the long term. A way has been suggested to complement the policy to make the entropy increased effectively and permanently.
	\end{abstract}
	
	\maketitle
	
	\section{Introduction}
	Recently, attempts have been made to carry out entropy-estimate based analysis of certain social scenarios \cite{Adnan R,Martin B,Thomas E}. An attempt was made in \cite{Pandey R K} to carry out such an entropic analysis of an simplified traffic intersection scenario. It was shown there that implementing traffic signals or constructing flyovers result in decrease of entropy thereby transforming the situation into a systematic behavior. It is expected that such estimates and analysis will help in understanding the social complexities in a better way through scientific and mathematical modelling \cite{Jacquemin A P,Matjaz P,Stephane L}. Such analysis might turn into tools for better assessment, analysis and predictions of social behavior and status \cite{Castellano C}. While analyzing these models on the basis of entropy estimates it is believed that processes that brings a positive change in the entropy of a social system would lead to equality whereas negative change will make the system function systematically but at the cost of equality \cite{Sam Overman E,Jost L,de Marchi}.\\
	
	The constitution of India \cite{constitution} has provisioned a Reservation Policy \cite{reservation} to ensure equality in the Indian society. This paper attempts to make a scientific analysis of the policy, based on the Entropy estimates of the society. Using the Entropic calculations, the effectiveness of this policy is ascertained scientifically. Attempt is made to verify that whether the social Entropy that is expected to increase due to this policy will meet the promise or not.\\
	
	The next section discusses the policy of reservation that is presently in operation in India. The need to have such a policy and the aim that it is looking to achieve are analyzed briefly. Using the same, a mathematical formulation is developed in the subsequent section to identify and estimate the Entropy of the society since the entire analysis is based on the changes that would be caused by the implementation of the policy. An example is then constructed with complete mathematical details in the next section to illustrate conclusively that the policy can increase the Entropy of the society in a short term only till the time the criteria to identify discriminated sections are not reutilized to reidentify the communities.
\section{The need for a Policy of Reservation}
After becoming independent in 1947, the government of India passed its own constitution that came into effect from January 26, 1949. The Constitution of India \cite{constitution} in its articles 15(4) and 16(4), provisioned a sound basis to implement a policy under affirmative action that was called - Reservation Policy \cite{reservation}. The policy makers realized that there are several communities in India that are socially discriminated against sharing positions associated with power and policy making. These ignored but large section of the society needed some affirmative action to get an equal opportunity as compared to those who were already better placed in the society. Based on this reality, a few categories were identified by grouping some communities together as Scheduled Castes (SC) and Scheduled Tribe (ST) at an initial stage and later some additional communities were grouped under Other Backward Class (OBC). These communities were identified by their caste identity and were made eligible for the provisions of affirmative action.\\

These categories were identified by estimating their share in the existing structure of the society on the scales of power and respect. Instead of economy the basis of identifying these categories were considered as per their say in the policy making, in drawing respect and power from the society.\\

The need for having this policy was felt because the share of these communities identified as SC, ST and OBC were not in the proportion to their share in the population. The policy since then, has intended to achieve proportionate representation of all these communities in the social structure of power and respect.\\

The policy was adopted after concluding correctly that since the share in the power structure was far lesser by a large number of communities despite having larger share in the population, they were automatically being discriminated by others who were in minority but were enjoying the advantage. It was realized in principle by the policy makers that to eliminate such practices of discrimination, all communities must get proportionate representation in the power structure. And thus, the policy was conceptualized to correct this skewed imbalance \cite{Maheshwari S R}. The structure of power and respect in the society gets automatically built up around the government jobs. It was also realized in principle, that government jobs dominantly contribute to the power structure and therefore the policy of reservation has been implemented in the government jobs belonging to all categories and stages. In particular, apart from those who could get into this social structure of respect and power without the help of the reservation policy, a minimum of 7.5\% from the ST community, a minimum of 15\% from SC and a minimum of 27\% from the OBC community have been assured of the share in this structure. Since these are not the upper caps, the policy looks at facilitating these communities in sharing more than these percentages, if the policy is successfully implemented for a reasonably long duration.
 \section{The Mathematical Model}
The present paper utilizes the idea of Social Entropy to carry out the analysis. As per the definition and the understanding of Entropy, any change to achieve equality will necessarily be associated with increase in Entropy. An attempt is made in this paper to do the Entropic analysis of this policy. It is well known that under the given conditions, the Social Entropy will be maximum when each community will get representation in the power structure in proportion to their share in the population of India. To build a model, let the Entropy of the society be defined in the following way.\\

Let us consider that there are total of $N$ positions available in the power structure of the society (loosely equal to the number of government jobs). As per the features of any power structure, let there be four sub-classes in this hierarchical structure having numbers as $N_1, N_2, N_3$ \& $N_4$\\ 
where
\begin{equation}\label{smn}
	\sum^{i=4}_{i=1} (N_i) = N
\end{equation}				
If the Social Power and Respect index of these positions are denoted by $Pow(N_i)$ then let them be chosen in the following order
\begin{equation}\label{hierarchy_pow}
	Pow(N_1) > Pow(N_2) > Pow(N_3) > Pow(N_4)
\end{equation}			
According to the characteristics of any hierarchical structure, this must be associated with
\begin{equation}\label{hierarchy_N}
	N_1 < N_2 < N_3 < N_4
\end{equation}

Now, let us try to formulate the identification Criteria of Communities to be selected for differential treatment to make them equal with others under the affirmative action strategy.\\

In general, people belonging to several castes will be occupying the N positions in some way. India has a large list consisting of several hundreds of castes identified in the government records. In the next step, let us try to identify some categories of people by grouping these castes, that are occupying these positions of power which is disproportionate to their share in the population. Let four such communities be identified by grouping these castes intelligently as $C_1, C_2, C_3$ \& $C_4$ that occupy these $N_1, N_2, N_3$ \& $N_4$ positions in a completely skewed manner as illustrated in the following observations.
\begin{enumerate}
	\item That $C_1, C_2, C_3$ \& $C_4$ have a\% b\% c\% \& d\% share in the population respectively.
	\item Let $a_{ij}$ denote the percentage share of the class $C_j$ in the positions $N_i$.
	\item That the percentage share in the most powerful positions denoted by $N_1$ is dominated by $C_1$ as compared to the other three $C_2, C_3$ \& $C_4$. (Mathematically, $a_{11}$ is very large as compared to $a_{12}, a_{13}$ \& $a_{14}$.)
	\item That the percentage share in the second most powerful positions denoted by $N_2$ is dominated by $C_2$ as compared to the other three $C_1, C_3$ \& $C_4$. (Mathematically, $a_{22}$ is very large as compared to $a_{21}, a_{23}$ \& $a_{24}$.)
	\item That the percentage share in the not so powerful positions denoted by $N_3$ is dominated by $C_3$ as compared to the other three $C_1, C_2$ \& $C_4$. (Mathematically, $a_{33}$ is very large as compared to $a_{31}, a_{32}$ \& $a_{34}$.)
	\item That the percentage share in the least powerful positions denoted by $N_4$ is dominated by $C_4$ as compared to the other three $C_1, C_2$ \& $C_3$. (Mathematically, $a_{44}$ is very large as compared to $a_{41}, a_{42}$ \& $a_{43}$.)
\end{enumerate}
To get the desirable proportionate representation in all the positions of power, the policy must result in changes in these parameters such that the following distribution is achieved:
\begin{equation}\label{maxent}
	a_{i1} : a_{i2} : a_{i3} : a_{i4} = a : b : c : d 	
\end{equation}	
for all $i=1,2,3,4$\\				
If the probability of $C_j$ occupying one of the positions in $N_i$ be given by $P_{ij}$, then 
\begin{equation}\label{probability}
	P_{ij} = N_{ij} / N	
\end{equation}								
Sixteen such parameters denoted by\\ $P_{11}, P_{12}, P_{13}, P_{14}, P_{21}, P_{22}, P_{23}, P_{24}, P_{31}, P_{32}$, $P_{33}, P_{34}, P_{41}, P_{42}, P_{43}$ \& $P_{44}$ will represent a particular distribution of people belonging to different categories occupying different positions of power. The Social Entropy \cite{Kenneth,Sargentis G F} of this distribution will then be given by
\begin{equation}\label{entropy}
	E = -\sum^{i=4}_{i=1} \sum^{j=4}_{j=1}( P_{ij} Log (P_{ij} ) )
\end{equation}

It can be easily shown that when the condition given in the equation (\ref{maxent}) is satisfied, the Entropy will attain its maximum value. In any skewed distribution therefore, the value of Entropy will be less than that and the success of the policy can be assessed by estimating how close the Entropy has reached to its maximum. 
The Policy of Reservation enforces proportionate representation in all the positions blindly irrespective of the hierarchy and thus promises to achieve proportionate representation in the entire power structure.\\

It is expected that in the new distribution, no community will be facing discrimination. But is this expectation realizable? Ironically, in an example discussed in the next section it is illustrated that when the proportionate distribution is attained, a new set of communities will start facing similar discrimination that was sought to be nullified through the policy. And so, with the identification of a new set of categories, the Entropy will be still far from reaching the desirable maximum value. The entire exercise therefore, becomes iso-entropic since it can never lead to a real increase in the Entropy.
\section{Illustrative Example}
To get the complete insight about this strategy, we discuss a possible scenario here, by illustrating a perfectly ordinary example.\\

Let us consider the following scenario. Let there be 400 positions of power and respect in a particular society and let these be distributed in four categories as
\begin{equation}\label{distribution_N}
	N_1 = 20, N_2 = 60, N_3 = 120, N_4 = 200 
\end{equation}

\begin{figure}
	\includegraphics[width=.8\linewidth]{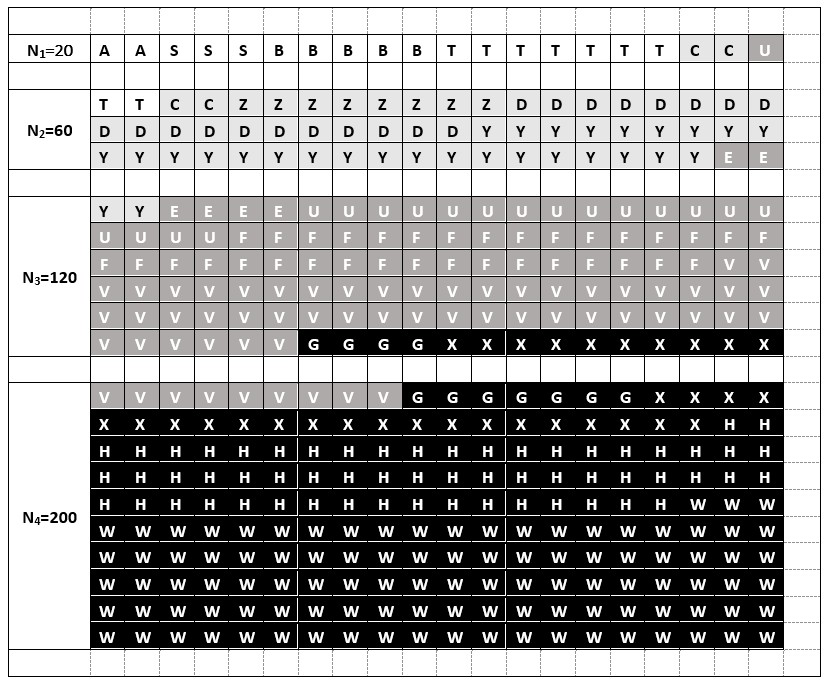}
	\caption {Initial given distribution of different 16 castes across 400 vacancies.\\
	\textbf{White Boxes:} Belong to $C_1$ Community consisting of $A, S, B, T$ castes\\
	\textbf{Light Grey Boxes:} Belong to $C_2$ community consisting of $C, Z, D, Y$ castes\\
	\textbf{Dark Grey Boxes:} Belong to $C_3$ community consisting of $E, U, F, V$ castes\\
	\textbf{Black Boxes:} Belong to $C_4$ community consisting of $G, X, H, W$ castes\\
	}
	\label{fig:distribution 1}
\end{figure}

Here $N_1$ belongs to the most powerful positions, $N_2$ belongs to the next powerful ones, $N_3$ are not so powerful but $N_4$ are the least powerful positions. Let there be sixteen communities (castes) of people, denoted by $A, B, C, D, E, F, G, H, S, T, U, V, W, X, Y, Z$, that are occupying these positions as shown in the Fig \ref{fig:distribution 1}. Using the criteria to identify communities for affirmative action it is noticed that 
\begin{enumerate}
	\item $A, S, B$ \& $T$ form the community $C_1$ that occupies the majority share 85\% in $N_1$.
	\item $C, Z, D$ \& $Y$ form the community $C_2$ that occupies the majority share 93\% in $N_2$.
	\item $E, U, F$ \& $V$ form the community $C_3$ that occupies the majority share 87\% in $N_3$.
	\item $G, X, H$ \& $W$ form the community $C_4$ that are left to occupy the majority share 96\% in $N_4$.
\end{enumerate}

This distribution is extremely skewed if compared with their population share as $C_1, C_2, C_3, C_4$ have 5\%, 15\%, 30\% and 50\% share in the population respectively.\\

\begin{figure}
	\includegraphics[width=.8\linewidth]{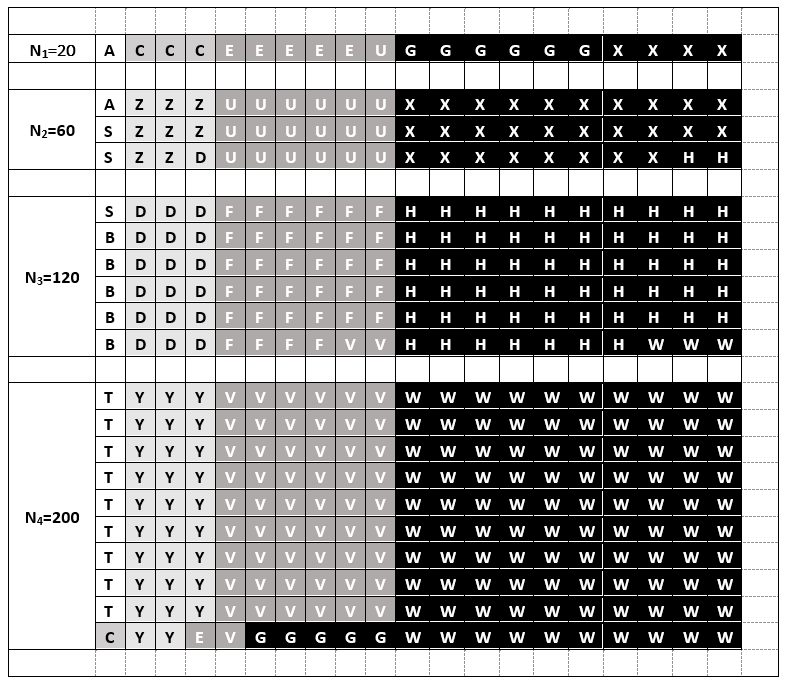}
	\caption {Distribution of different 16 castes across 400 vacancies after implementation of Reservation Policy.\\
		\textbf{White Boxes:} Belong to $C_1$ Community  consisting of $A, S, B, T$ castes\\
		\textbf{Light Grey Boxes:} Belong to $C_2$ community  consisting of $C, Z, D, Y$ castes\\
		\textbf{Dark Grey Boxes:} Belong to $C_3$ community  consisting of $E, U, F, V$ castes\\
		\textbf{Black Boxes:} Belong to $C_4$ community  consisting of $G, X, H, W$ castes\\
	}
	\label{fig:distribution 2}
\end{figure}

The Entropy of this distribution is estimated in the Table \ref{Table one} using the equation( \ref{probability}) and equation(\ref{entropy}) as 0.7278. When the Policy of reservation is enforced, all the categories $C_1, C_2, C_3, C_4$ will get proportionate representation separately in all stages of power. Therefore, we expect a distribution that is shown in Fig \ref{fig:distribution 2}. The Entropy associated with this distribution is estimated as 0.9870 for such a distribution as shown in the Table \ref{Table one}. This value is apparently very close to its maximum and therefore illustrates the success of the policy in achieving a situation wherein there is no discrimination.\\ 

To our surprise however, there is a twist in this understanding. If the criteria discussed in the earlier section is applied in this distribution as displayed in Fig \ref{fig:distribution 2}, one can identify four new categories by grouping the communities differently. To illustrate this, the following new categories are now identified with the features mentioned therein:
\begin{enumerate}
	\item $A, C, E$ \& $G$ form the community $C^{'}_1$ that occupies the majority share 75\% in $N_1$.
	\item $S, Z, U$ \& $X$ form the community $C^{'}_2$ that occupies the majority share 93\% in $N_2$.
	\item $B, D, F$ \& $H$ form the community $C^{'}_3$ that occupies the majority share 95\% in $N_3$.
	\item $T, U, V$ \& $W$ form the community $C^{'}_4$ left to occupy the majority share 96\% in $N_4$.
\end{enumerate}

\begin{figure}
	\includegraphics[width=.8\linewidth]{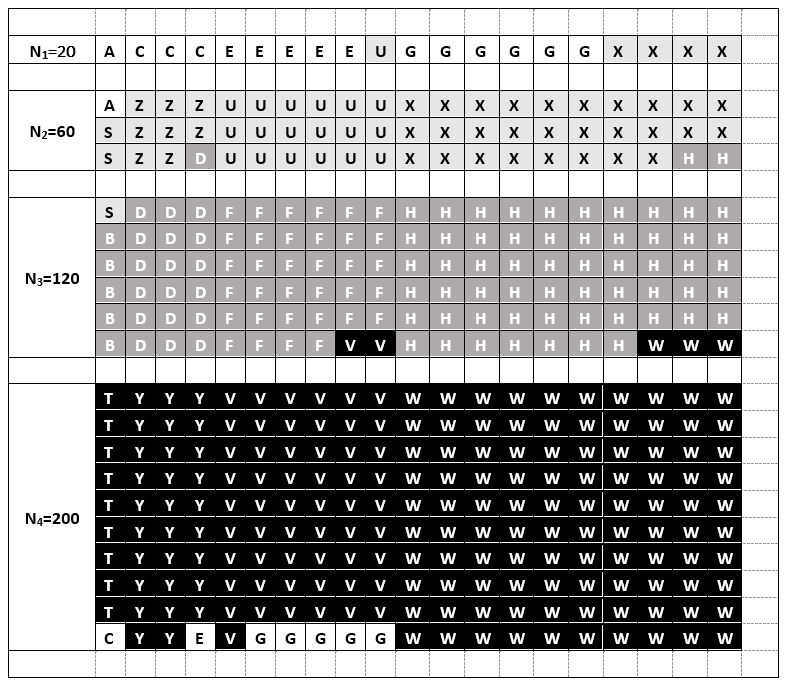}
	\caption {Distribution of different 16 castes across 400 vacancies after regrouping castes to form the four Communities according to the criteria decided initially.\\
		\textbf{White Boxes:} Belong to $C^{'}_1$ Community consisting of $A, C, E, G$ castes\\
		\textbf{Light Grey Boxes:} Belong to $C^{'}_2$ community consisting of $S, Z, U, X$ castes\\
		\textbf{Dark Grey Boxes:} Belong to $C^{'}_3$ community consisting of $B, D, F, H$ castes\\
		\textbf{Black Boxes:} Belong to $C^{'}_4$ community consisting of $T, Y, V, W$ castes\\
	}
	\label{fig:distribution 3}
\end{figure}

We see in the Fig \ref{fig:distribution 3} wherein the same distribution that exists in Fig \ref{fig:distribution 2}, becomes skewed with the new categories $C^{'}_1, C^{'}_2, C^{'}_3, C^{'}_4$ ($C^{'}_4$ now being the most discriminated set of communities). It is shown in the Table \ref{Table one} that Entropy of this distribution becomes 0.5884, showing no improvement on the scale of equality from its earlier value of 0.6238. Obviously, the increase in Entropy noticed earlier, was only as we were ignoring and delaying to acknowledge the change in power structure that the policy had brought in the society. The moment, the changes were acknowledged in Fig \ref{fig:distribution 3}, the policy is realized as iso-entropic, promising no improvement on the scale of equality.
	\section{Conclusion and Solution}
	The policy apparently promises to achieve the proportionate representation of all the categories in all the positions to ensure that no category of people is left out in this sharing of power and respect in the society. However, a careful analysis as done in the previous section proves that the success lies in having ignored the new communities that will start satisfying the criteria for being recognized as enjoying power and/or are beginning to being denied their due share in the power structure. The moment the criteria is used for regrouping the communities, Entropy will fail to show any appreciable increase. And thus, equality will never be achieved.\\
	
	Let us try to find out the factor that is again bringing the Entropy back to its initial levels. In fact, since the power associated with $N_1, N_2, N_3$ \& $N_4$ are hierarchically ordered, the share of communities in these categories are bound to be skewedly different. Moreover, since only those who would have larger share of presence in $N_1$ (smallest of all) can be considered powerful, leads to a situation wherein a large number of communities will start feeling ignored.\\ 
	
	The solution therefore exists in bridging the difference between the power sub-structures. For example, if the difference between the power and respect associated with $N_2$ \& $N_3$ gets eliminated we can expect similar share in these positions by all the communities. This is reflected in the Table \ref{Table one} by proportionately distributing the numbers between $N_2$ \& $N_3$. Entropy of this distribution wherein the parity between $N_2$ \& $N_3$ is restored (see Appendix), is estimated as 0.7278 in the Table \ref{Table one} (under Case 4) which is a definite improvement over the distribution of Fig \ref{fig:distribution 1}. Moreover, when proportionate distribution is achieved by reducing/eliminating the difference between the power sub-structures, the revisiting of community-grouping will make no difference in the final estimate of the Entropy.\\
	\begin{center}
	\begin{table}[!htbp]
		\begin{tabular}{ | m{1cm} | m{1cm} | m{1cm} | m{1.4cm} | m{1cm} | m{1.4cm} | m{1cm} |m{1.4cm} | m{1cm} |m{1.4cm} | m{1cm} | m{1.4cm} | }
			\hline
			$i$ & $j$ & $N_{ij}$ & $P_{ij}$ & $N_{ij}$ & $P_{ij}$ & $N_{ij}$ & $P_{ij}$ & $N_{ij}$ & $P_{ij}$ & $N_{ij}$ & $P_{ij}$ \\ 
			\hline
			&& \textbf{Case}&1 & \textbf{Case}& 2 & \textbf{Case}& 3 & \textbf{Case}& 4 & \textbf{Case}& 5 \\
			\hline
			1 & 1 & 17 & 0.0425 & 1 & 0.0025 & 15 & 0.0375 & 17 & 0.0425& 17 & 0.0425 \\ 
			\hline
			1 & 2 & 2 & 0.005&	3&	0.0075&	5&	0.0125&	2&	0.005&	2&	0.005 \\ 
			\hline
			1 & 3 & 1&	0.0025&	6&	0.015&	0&	0&	1&	0.0025&	1&	0.0025\\ 
			\hline
			1 & 4 & 0&	0&	10&	0.025&	0&	0&	0&	0&	0&	0 \\ 
			\hline
			2 & 1 & 2&	0.005&	3&	0.0075&	1&	0.0025&	1&	0.0025&	2&	0.005 \\ 
			\hline
			2 & 2 & 56&	0.14&	9&	0.0225&	56&	0.14&	19&	0.0475&	56&	0.14\\ 
			\hline
			2 & 3 & 2&	0.005&	18&	0.045&	3&	0.0075&	35&	0.0875&	2&	0.005 \\  
			\hline
			2 & 4 & 0&	0&	30&	0.075&	0&	0&	5&	0.0125&	0&	0 \\  
			\hline
			3 & 1 & 0&	0&	6&	0.015&	0&	0&	1&	0.0025&	0&	0\\  
			\hline
			3 & 2 & 2&	0.005&	18&	0.045&	1&	0.0025&	39&	0.0975&	2&	0.005 \\  
			\hline
			3 & 3 & 104&	0.26&	36&	0.09&	114&	0.285&	71&	0.1775&	42&	0.105 \\  
			\hline
			3 & 4 & 14&	0.035&	60&	0.15&	5&	0.0125&	9&	0.0225&	77&	0.1925 \\  
			\hline
			4 & 1 & 0&	0&	9&	0.0225&	7&	0.0175&	0&	0&	0&	0 \\  
			\hline
			4 & 2 & 0&	0&	30&	0.075&	0&	0&	0&	0&	0&	0 \\  
			\hline
			4 & 3 & 9&	0.0225&	56&	0.14&	0&	0&	9&	0.0225&	71&	0.1775 \\  
			\hline
			4 & 4 & 191&	0.4775&	105&	0.2625&	193&	0.4825&	191&	0.4775&	128&	0.32 \\  
			\hline
			\multicolumn{3}{|c|}{\textbf{Entropy}}
			& 0.6238 & & 0.9870 & & 0.5884 & & 0.7278 & & 0.7625 \\
			\hline
		\end{tabular}
		\caption{ Entries in the columns of this table are as follows.\\ \textbf{Case 1:} Deals with the distribution shown in the Fig \ref{fig:distribution 1}\\ \textbf{Case 2:} Deals with the distribution shown in the Fig \ref{fig:distribution 2}\\ \textbf{Case 3:} Deals with the distribution shown in the Fig \ref{fig:distribution 3}\\ \textbf{Case 4:} Deals with the distribution given in Fig \ref{fig:distribution 1} is modified on restoration of parity between $N_2$ and $N_3$\\ \textbf{Case 5:} Deals with the distribution given in Fig \ref{fig:distribution 1} is modified on restoration of parity between $N_3$ and $N_4$}
		\label{Table one}
	\end{table}
\end{center}

	Due to long duration of sufferings caused by slavery and cultural invasions probably, the Indian society lost the sense to acknowledge dignity of labour and work. An urge to change one’s profession merely to earn respect in the society, originates from our collective loss of sense in respecting dignity of work. Unless this sense is restored, the Reservation Policy, in itself, will not be enough to achieve the aim of ensuring equality in the society. One of the ways to restore this sense could be to reduce the gap in the income associated with the stages denoted by $N_1, N_2, N_3$ \& $N_4$ so that these positions do not get treated differently at least due to the financial status associated with these positions. As reflected in the Table \ref{Table one}, even a restoration of parity between $N_2$ \& $N_3$ has been able to improve the Entropy from 0.6238 to 0.7278 that is closer to its maximum value of 0.9870. Just to help in choosing a strategy one must note that, restoring parity between $N_3$ \& $N_4$ (see Appendix), instead of between $N_2$ \& $N_3$, shows the promise of improvement in the Entropy to 0.7625.
\section{Appendix}
To restore the parity between $N_2$ and $N_3$ the calculation is done in the following way: 
If a community $C_i$ has a combined strength of $N_c$ then it is redistributed among $N_2$ and $N_3$ in the ratio 60:120 (or1:2) since $N_2=60$ and $N_3=120$. According to the Fig \ref{fig:distribution 1}, community $C_1, C_2, C_3$ and $C_4$ has a combined strength as 2, 58, 116, 14 respectively. These are then proportionately distributed in $N_2$ and $N_3$ in the ratio 1:2 as given in the Table \ref{Table one}. The two levels given by $N_2$ and $N_3$ are not made one to keep the system comparable with the initial state when the system has four levels of hierarchy.\\

Similarly, to restore the parity between $N_3$ and $N_4$ following calculation is done: 
According to the Fig \ref{fig:distribution 1}, community $C_1, C_2, C_3$ and $C_4$ has a combined strength as 0, 2, 113, 205 respectively. These are then proportionately distributed in $N_3$ and $N_4$ in the ratio 120:200 (or 3:5) as given in the Table \ref{Table one}. 

\section{acknowledgement}
\begin{acknowledgments}
	The authors thank Kirori Mal College and Hindu College to have provided us the required environment to carry out this work.
\end{acknowledgments}

\end{document}